\documentclass[runningheads,a4paper]{llncs}
\usepackage{amssymb}
\usepackage{graphicx}
\usepackage{indentfirst}
\usepackage{url}
\urldef{\mailsa}\path|{alfred.hofmann, ursula.barth, ingrid.haas, frank.holzwarth,|
\urldef{\mailsb}\path|anna.kramer, leonie.kunz, christine.reiss, nicole.sator,|
\urldef{\mailsc}\path|erika.siebert-cole, peter.strasser, lncs}@springer.com|
\newcommand{\keywords}[1]{\par\addvspace\baselineskip
\noindent\keywordname\enspace\ignorespaces#1}
\begin{document}

\mainmatter  

\title{A quantum algorithm for the dihedral hidden subgroup problem based on algorithm SV}

\titlerunning{}

%
%
\author{Li Fada,
\ Bao Wansu
\thanks{2010thzz@sina.com},
\ Fu Xiangqun}
\authorrunning{}

\institute{the PLA Information Engineering University, Zhengzhou 450004, China}

%
%

\toctitle{}
\tocauthor{}
\maketitle

\begin{abstract}
To accelerate the algorithms for the dihedral hidden subgroup problem, we present a new algorithm based on algorithm SV. A subroutine is given to get a transition quantum state by constructing a phase filter function, then the measurement basis are derived based on the algorithm SV for solving low density subset problem. Finally, the parity of slope $s$ is revealed by the measurement. This algorithm takes $O(n)$ quantum space and $O({n^2})$ classical space, which is superior to existing algorithms, for a relatively small $n$ (about $n \le 6400$ ), it takes $\sqrt n {(\log \max {a_{ij}})^3}$ computation time, improving upon the previous time complexity of ${2^{O(\sqrt n )}}$.
\end{abstract}
\section{Introduction}
Quantum computers are believed to solve certain problems much more efficiently than any classical machine. As a result, they dramatically improve the decoding ability to public key cryptography. For example, the shor algorithm \cite{1} can break the RSA and ECC in polynomial time on a quantum computer. Correspondingly quantum algorithms are known as important tools to break public key cryptography in the future.

It has been proved that the integer factorization problem can be reduced to the abelian hidden subgroup problem. In fact, most of the mathematic problems that quantum algorithms can solve effectively are belong to abelian hidden subgroup problems, but there are no known algorithms for non-abelian hidden subgroup problem in polynomial time. In addition, the post quantum public key cryptography \cite{2,3,4,5} are based on some mathematical problems which are closely related to non-abelian hidden subgroup problem. Thus the research on algorithms for non-abelian hidden subgroup problem is important for the security of post quantum public key cryptography.

Among the post quantum public key cryptography, lattice-based cryptography has a fast development. Since some lattice problems such as unique Shortest Vector Problem(uSVP) can be used to design the trapdoor one-way function, many public key encryption schemes based on lattice problems\cite{6,7} have been proposed. However, uSVP can be reduced to dihedral hidden subgroup problem which is a kind of non-abelian hidden subgroup problem\cite{8}, therefore the study on quantum algorithm for the dihedral hidden subgroup problem has great significance for the security of lattice-based cryptography.

M.Ettingert and P.Hoyer\cite{9} proposed the first quantum algorithm for the dihedral hidden subgroup problem, they reduced the hidden subgroup problem over ${D_{{2^n}}}$ to finding the slope $s$ of the hidden subgroup $H =  < y{x^s} > $, the time complexity is $O({2^n})$. G.Kuperberg\cite{10} further reduced this problem to finding the parity of the slope $s$. He constructed a sieve that began with ${2^{O(\sqrt n )}}$ quantum states $\left| {{\varphi _k}} \right\rangle  \propto \left| 0 \right\rangle  + {{\mathop{\rm e}\nolimits} ^{(2\pi iks/N)}}\left| 1 \right\rangle $ (The symbol $ \propto $ means ¡°proportional to¡±, so that the normalization and global phase can be omitted),  after an exhaustive search at each stage of the sieve, he found two quantum states $\left| {{\varphi _k}} \right\rangle $ and $\left| {{\varphi _l}} \right\rangle $ such that $k$ and $l$ agreed in $O(\sqrt n )$ low bits, then the new quantum state $\left| {{\varphi _{k'}}} \right\rangle $ could be produced such that $O(\sqrt n )$ low bits of $k'$ is 0, at last it could produce the target quantum state $\left| 0 \right\rangle  + {( - 1)^s}\left| 1 \right\rangle $ which revealed the parity of $s$ by the measurement. The time complexity is ${2^{O(\sqrt n )}}$ and it also requires ${2^{O(\sqrt n )}}$ quantum space. O.Redev\cite{11} gave another algorithm, he produced the quantum state $\sum\limits_{j = 1}^m {{e^{2\pi is a ({x_j})/N}}\left| {{x_j}} \right\rangle }$ ( the phase function $a({x_j})$ is a linear function with respect to ${x_j}$), after an exhaustive search for the solutions to a congruence equation respect to $a ({x_j})$, he coudl get the quantum states that $a ({x_j})$ agreed in $O(\sqrt n )$ low bits. The algorithm requires ${2^{O(\sqrt {n\log n} )}}$ computation time but only $O(\sqrt {n\log n} ) $ quantum space. G. Kuperberg\cite{12} then gave a new algorithm, he obtained $\sum\limits_{1 \le {x_1} < {l_1},1 \le {x_2} < {l_2}}^{} {{e^{2\pi is({a_1}({x_1}) + {a_2}({x_2}))/N}}\left| {{x_1},{x_2}} \right\rangle }$ by preprocessing $\sum\limits_{j = 1}^m {{e^{2\pi is a({x_j})/N}}\left| {{x_j}} \right\rangle } $ and established the truth table of ${a_j}({x_j})\bmod {2^m}$ to find the quantum state which ${a _1}({x_1}) + {a_2}({x_2})$ agreed in $O(\sqrt n )$ low bits. The algorithm takes ${2^{O(\sqrt n )}}$ computation time and only $O(\sqrt n )$ quantum space, but needs ${2^{O(\sqrt n )}}$ classical space. This is the best algorithm for dihedral hidden subgroup problem so far. However, this algorithm also uses the method of pairing and iteration, large classical space is required to store the phase of quantum states waiting for pairing. How to design a new quantum algorithm to decrease the time and space complexity still needs further study.

The Lenstra Lenstra Lovasz lattice reduction algorithm (LLL algorithm) was proposed in 1982, it was recognized as one of the most important mathematical tool because of its useful applicability. Among the most famous applications are those in lattice theory, that is, the shortest vector problem (SVP) and the closest vector problem (CVP), another important application is to solve low density subset sum problem, J.C.Lagarias and A.M.Odlyzko reduced this problem to searching for a particular short vector in a lattice\cite{15}. Thus it is necessary to investigate the application of LLL algorithm on the dihedral hidden subgroup problem.

The following is the main theorem of this paper. we will prove this theorem in section 3.
\\\textbf{Theorem 1.1} If there exists a solution to a low density subset sum problem, then there exists a quantum algorithm that solves the dihedral hidden subgroup problem.

We give a quantum algorithm for the dihedral hidden subgroup problem based on algorithm SV. Section 3.1 and section 3.2 can be regarded as some quantum processing that reduces the dihedral hidden subgroup problem to a low density subset sum problem. Thus the quantum algorithm we present can be seen as another application of the LLL algorithm. In section 3.1, we design a transition quantum state generating algorithm by constructing a phase filter function, in section 3.2, we convert the problem of finding the proper measurement basis to a low subset sum problem in $\sqrt n $ dimension which can be solved by algorithm SV. When $\sqrt n $ is small, algorithm SV will success with high probability, then perform a measurement on the transition quantum state in the basis obtained to get the target quantum state. At last, the parity of $s$ can be revealed by measurement on the target quantum state in $\left|  \pm  \right\rangle $ basis. The algorithm decreases time and space complexity obviously (especially in practice) by comparison with the results in \cite{13} (which takes ${2^{O(\sqrt n )}}$ computation time, $O(\sqrt n )$ quantum space and ${2^{O(\sqrt n )}}$ classical space).
\section{Preliminaries}
\textbf{Definition 2.1 (dihedral group)} the dihedral group ${D_N}$ with $2N$ elements is a symmetry group of a regular N-gon in the plane has the conventional presentation
$${D_N} =  < x,y\left| {{x^N} = {y^2}} \right. = yxyx = 1 > $$

An element of the form ${x^s}$ is a rotation and an element of the form $y{x^s}$ is a reflection. The parameter $s$ is the slope of the reflection $y{x^s}$. We can describe ${D_N}$ as a semi-direct product ${D_N} = {Z_N} \times {}_\phi {Z_2}$ where $(a,b)$ represents the element ${x^a}{y^b}$, so the law of this semi-direct product is given by
$$({a_1},{b_1})({a_2},{b_2}) = ({a_1} + {( - 1)^{{b_1}}}{a_2},{b_1} + {b_2})$$

\textbf{Definition 2.2 (DHSP)} Let $f$ be a function from a dihedral group ${D_N}$ to a finite set $X$ such that $f$ is constant on the cosets of a subgroup $H$, and distinct on each coset. Given a quantum black box for performing the unitary transform
$${U_f}\left| {(a,b)} \right\rangle \left| h \right\rangle  = \left| {(a,b)} \right\rangle \left| {h \oplus f(a,b)} \right\rangle $$

For $(a,b) \in {D_N},h \in H$ and $ \oplus $ an appropriately chosen binary operation on X, find a generating set for $H$.

\textbf{Proposition 2.1\cite{6}} Finding an arbitrary hidden subgroup $H$ of ${D_N}$ reduces to finding the slope of a hidden reflection.

Kuperberg\cite{10} reduced DHSP to find the parity of the slope of a hidden reflection further. In this paper we also design the quantum algorithm with this method.

\textbf{Definition 2.3 (subset sum problem)} Given positive integers ${a_i}$ and $c$, the subset sum problem is to find variables ${x_i} \in \{ 0,1\} $, such that
$$\sum\limits_{i = 1}^m {{a_i}{x_i}}  = c$$

The density of such problem is defined as follows:
$$d = \frac{m}{{{{\log }_2}\mathop {\max }\limits_{1 \le i \le m} {a_i}}}$$

J.C.Lagarias and A.M.Odlyzko proposed the algorithm SV to solve low density subset sum problem, this algorithm finds the solution with high probability when $d < (2 - \delta ){({\log _2}4/3)^{ - 1}}{m^{ - 1}}$ for any fixed $\delta$.
\section{Reduction from dihedral hidden subgroup problem to low density subset sum problem}
In this section we will prove Theorem 1.1. We begin by presenting a transition quantum state generating algorithm in section 3.1, we then reduce the problem of finding the measurement basis to a low density subset sum problem in section 3.2. A measurement on the transition quantum state in the corresponding basis reveals the target quantum state.

\textbf{3.1 Transition quantum state generating algorithm}

For simplicity, we describe the dihedral hidden subgroup problem with $N = {2^n}$, the key step of the algorithm is to find the target quantum state $\left| 0 \right\rangle  + {( - 1)^s}\left| 1 \right\rangle$. Exhaustive search costs exponential time, thus the method of pairing and iteration was used in \cite{10,11,12} so that one can get the target quantum state gradually. However this method needs large classical space for storing the phase of quantum states waiting for pairing. In this section, we give a transition quantum state generating algorithm, rather than using the method of pairing and iteration, we calibrate the phase of preprocessed quantum state to obtain a transition quantum state $\left| {{x^{(1)}}} \right\rangle  + {( - 1)^s}\left| {{x^{(2)}}} \right\rangle  + {( - 1)^{ts}}\left| {{x^{(3)}}} \right\rangle ...$ which is very close to the target state.

To finish the algorithm, we need to preprocess the initial quantum state $\left| 0 \right\rangle $ in the data register.

\textbf{3.1.1 preprocessing of the initial quantum state}

First, we use the quantum fourier sampling\cite{13} to get the quantum state below
$$\left| {{\varphi _k}} \right\rangle  \propto \left| 0 \right\rangle  + {{\mathop{\rm e}\nolimits} ^{(2\pi iks/N)}}\left| 1 \right\rangle $$

Notice that as we sample a quantum state above, the index $k \in \{ 0,1,{...2^n} - 1\} $ is also fixed and stored into classical memory.

Then, we give the tensor product of the above $m = \left\lceil {\sqrt n } \right\rceil $ quantum states $\left| {{\varphi _{{a_{11}}}}} \right\rangle ,\left| {{\varphi _{{a_{12}}}}} \right\rangle ...,\left| {{\varphi _{{a_{1m}}}}} \right\rangle $ and obtain
$$\left| \varphi  \right\rangle  = \left| {{\varphi _{{a_{11}}}}} \right\rangle  \otimes \left| {{\varphi _{{a_{12}}}}} \right\rangle  \otimes ... \otimes \left| {{\varphi _{{a_{1m}}}}} \right\rangle $$

Define $ a_1 = ({a_{11}},...,{a_{1m}})$,  the binary representation of $x_1$ is $({x_{11}},...,{x_{1m}})$, then we have
$$\left| \varphi  \right\rangle  = \sum\limits_{{x_1} \in \{ 0,...,{2^m} - 1\} } {\exp (2\pi is{a_1}({x_1})/{2^n})\left| {{x_1}} \right\rangle } $$

Where ${a_1}({x_1}) = \sum\limits_{i = 1}^m {{a_{1i}}{x_{1i}}} $. For simplicity, we assume ${m^2} = n$. To ensure the success probability, tensor together $m = \left\lceil {\sqrt n } \right\rceil $ quantum states of the form (1) to obtain
$$\sum\limits_{} {\exp (2\pi i({a_1}({x_1}) + {a_2}({x_2}) + ...{a_m}({x_m}))s/{2^n})} \left| {{x_1},{x_2},...,{x_m}} \right\rangle$$

As the narrative convenience, we call ${a_1}({x_1}) + {a_2}({x_2}) + ...{a_m}({x_m})$ the phase function of $\left| {{x_1},{x_2},...,{x_m}} \right\rangle$, we only consider ${a_1}({x_1}) + {a_2}({x_2}) + ...{a_m}({x_m})\bmod {2^n}$, that is, the last $n$ bits information in the binary representation of the value.

\textbf{3.1.2 Transition quantum state generating algorithm}

The basic method of the transition quantum state generating algorithm is to produce a superposition, the ground states $\left| {{x_1},{x_2},...,{x_m}} \right\rangle$ of the superposition meet the requirement that their last $n - 1$ bits of phase function are the same. Specifically, we adds $m$ target registers and compute the phase filter function ${g_i}({x_i}) \equiv {a_i}{x_i}\bmod {2^{n - 1}}$ in each register and then measure these $m$ registers to get ${c_1},...,{c_m}$, that is, ${g_i}({x_i}) = {c_i}$. Meanwhile the quantum state in the data register collapses to the designated superposition. We can get the transition quantum state after eliminating a global phase. The complete procedure is as follows:

\textbf{Input} (1) one $n$ qubits data register initialized to $\left| 0 \right\rangle $; (2) $m$ $n$ qubits target registers initialized to $\left| 0 \right\rangle $; (3) a black box which performs the operation ${U_i}\left| {{x_i}} \right\rangle \left| y \right\rangle  = \left| {{x_i}} \right\rangle \left| {y \oplus {g_i}({x_i})} \right\rangle $, for ${g_i}({x_i}) \equiv {a_i}{x_i}\bmod {2^{n - 1}}$, where $1 \le i \le m$.

\textbf{Output} the transition quantum state $\left| {{x^{(1)}}} \right\rangle  + {( - 1)^s}\left| {{x^{(2)}}} \right\rangle  + {( - 1)^{ts}}\left| {{x^{(3)}}} \right\rangle ...$

\textbf{Procedure}

Step1 Preprocess the initial quantum state$\left| 0 \right\rangle $ in the data register as described in 3.1 and obtain
$$\sum\limits_{} {\exp (2\pi i({a_1}({x_1}) + {a_2}({x_2}) + ...{a_m}({x_m}))s/{2^n})} \left| {{x_1},{x_2},...,{x_m}} \right\rangle \left| 0 \right\rangle ...\left| 0 \right\rangle$$

Step2 Apply the black box to performs the operation ${U_i}\left| {{x_i}} \right\rangle \left| y \right\rangle  = \left| {{x_i}} \right\rangle \left| {y \oplus {g_i}({x_i})} \right\rangle $ and obtain
$$\sum\limits_{} {\exp (2\pi i({a_1}({x_1}) + {a_2}({x_2}) + ...{a_m}({x_m}))s/{2^n})} \left| {{x_1},{x_2},...,{x_m}} \right\rangle \left| {{g_1}({x_1})} \right\rangle ...\left| {{g_m}({x_m})} \right\rangle $$

Step3 Measure the last $m$ target registers, assume we get the results ${c_1},...,{c_m}$, set $c = {c_1} + ... + {c_m}\bmod {2^{n - 1}}$ and store it into classical memory. The quantum state in data register collapses to
$$\sum\limits_{a(x) = c\bmod {2^{n - 1}}} {\exp (2\pi ia(x)s/{2^n})} \left| x \right\rangle $$

Where $a(x) = {a_1}({x_1}) + {a_2}({x_2}) + ...{a_m}({x_m})$, $x = ({x_1},...{x_m})$, and ${x^{(1)}},{x^{(2)}},...$ denote $x$ that satisfies $a(x) = c\bmod {2^{n - 1}}$.

Notice that the number of $x$ that satisfies $a(x) = c\bmod {2^{n - 1}}$ is more than 2 with high probability (see Theorem 6.1). At this moment, the value of phase function $a({x^{(i)}})$ are the same in last $n - 1$ bits, that is $a({x^{(i)}}) = 0 + c$ or ${2^{n - 1}} + c$, we can get $a({x^{(2)}}) - a({x^{(1)}}) = {2^{n - 1}}$ with probabilities and thus get the transition quantum state $\left| {{x^{(1)}}} \right\rangle  + {( - 1)^s}\left| {{x^{(2)}}} \right\rangle  + {( - 1)^{ts}}\left| {{x^{(3)}}} \right\rangle ...$ by eliminating global phase factor $\exp (2\pi ia({x^{(1)}})s/{2^n})$.

\textbf{3.2 Obtaining target quantum state}

As mentioned above, getting the parity of $s$ requires the corresponding measurement on the transition quantum state, in this section we reduce the problem of finding the measurement basis to a low density subset sum problem, then by calling the algorithm SV we can get the measurement basis.

We begin with the following two simple claims

\textbf{Claim 3.1\cite{11}} For $\sum\limits_{j = 1}^l {e({\phi _j})\left| {{x^{(j)}}} \right\rangle } $, if the ground states $\left| {{x^{(1)}}} \right\rangle $ and $\left| {{x^{(2)}}} \right\rangle $ are known, then we can perform a projective measurement on the subspace spanned by $\left| {{x^{(1)}}} \right\rangle $ and $\left| {{x^{(2)}}} \right\rangle $ to get the state $e({\phi _1})\left| {{x^{(1)}}} \right\rangle  + e({\phi _2})\left| {{x^{(2)}}} \right\rangle $ with constant probability.

\textbf{Claim 3.2\cite{8}} For any two basis states $\left| {{x^{(1)}}} \right\rangle $ and$\left| {{x^{(2)}}} \right\rangle $,${x^{(1)}} \ne {x^{(2)}}$, there exists a routine such that given the state $\left| {{x^{(1)}}} \right\rangle  + e(\phi )\left| {{x^{(2)}}} \right\rangle $ outputs the state$\left| 0 \right\rangle  + e(\phi )\left| 1 \right\rangle $.

These two claims give the procedure of getting the target quantum state by the measurement in $\left| {{x^{(1)}}} \right\rangle ,\left| {{x^{(2)}}} \right\rangle $ basis, and then we can get the parity of $s$ by the further measurement on target quantum state in $\left|  \pm  \right\rangle $ basis. Next we use the special property of ${x^{(i)}},i = 1,2$ which satisfies $a({x^{(j)}}) \equiv c\bmod {2^{n - 1}}$ to prove that getting the value of ${x^{(i)}},i = 1,2$ is equivalent to solving a low density subset sum problem, then we call the algorithm SV \cite{15} to get ${x^{(i)}},i = 1,2$.
${x^{(i)}},i = 1,2$ satisfy $a({x^{(i)}}) \equiv c\bmod {2^{n - 1}}$,that is
$$\left\{ \begin{array}{l}
{a_1}{x_1}^{(i)} \equiv {c_1}\bmod {2^{n - 1}}\\
{a_2}{x_2}^{(i)} \equiv {c_2}\bmod {2^{n - 1}}\\
 \vdots \\
{a_m}{x_m}^{(i)} \equiv {c_m}\bmod {2^{n - 1}}
\end{array} \right.$$

Thus, we decompose the procedure of obtaining ${x^{(i)}}$ into obtaining there $m$ components ${x_j}^{(i)}$, since the $m$ equations have the same structure, we only solve the following equation to compute ${x_1}^{(i)}$.
$${a_{11}}{x_{11}}^{(i)} + {a_{12}}{x_{12}}^{(i)} + ... + {a_{1m}}{x_{1m}}^{(i)} \equiv {c_1}\bmod {2^{n - 1}}$$

Since ${a_{ji}} < {2^n}$, we can know that $\mathop {\max }\limits_{{x_1} \in {{\{ 0,1\} }^m}} {a_1}({x_1}^{(i)}) \le m{2^n}$, thus the equation above can be written as follows
$${a_{11}}{x_{11}}^{(i)} + {a_{12}}{x_{12}}^{(i)} + ... + {a_{1m}}{x_{1m}}^{(i)} = {c_1} + t{2^{n - 1}},0 \le t < 2m   \eqno{(2)} $$

We consider the solutions of the above equation when $t \in \{ 0,...2m - 1\} $. We consider if there is ${x_1}^{(i)}$ satisfy the equation when $t = 0$. Since $P(\mathop {\max }\limits_{1 \le l \le m} {a_{1l}} \ge {2^{n - 1}}) = 1 - {(\frac{1}{2})^m} \to 1$, the density of this subset sum problem is $d < \frac{m}{{{{\log }_2}({2^{n - 1}})}}$, thus the problem is reduced to solve the subset sum problem with size $\sqrt n $ and density $d < \frac{1}{{\sqrt n }}$. Then, we can call the algorithm SV until $t = 2m - 1$ to get the solutions and substitute them into equation to verify their validity.

Repeat the above procedure $m$ times to get $({x_1}^{(i)},...,{x_m}^{(i)})$, select two of them to be ${x}^{(i)},i \in \{ 1,2\} $, or else the algorithm fails. If we know the value of ${x}^{(i)},i \in \{ 1,2\} $, we can we can prepare the measurement basis and obtain the target state $\left| 0 \right\rangle  + {( - 1)^s}\left| 1 \right\rangle $ according to claim 3.1 and claim 3.2.

This concludes the proof of theorem 1.1.

\section{Quantum algorithm for the dihedral hidden subgroup problem based on algorithm SV}
In this section we combine the ideas of sections 3 and the algorithm SV to make a complete algorithm for the dihedral hidden subgroup problem.

\textbf{Input} (1) two  qubits data registers initialized to $\left| 0 \right\rangle $; (2) $m + 1$ $n$ qubits target registers initialized to $\left| 0 \right\rangle $; (3) a black box which performs the operation $U\left| x \right\rangle \left| y \right\rangle  = \left| x \right\rangle \left| {y \oplus f(x)} \right\rangle $, for $f(x)$ defined in definition 2.2 ;(4) a black box which performs the operation ${U_i}\left| {{x_i}} \right\rangle \left| y \right\rangle  = \left| {{x_i}} \right\rangle \left| {y \oplus {g_i}({x_i})} \right\rangle $, for ${g_i}({x_i}) \equiv {a_i}{x_i}\bmod {2^{n - 1}}$, where $1 \le i \le m$.$n$

\textbf{Output} the parity of $s$.

\textbf{Procedure}

Step1 Use the quantum fourier sampling to get $n$ quantum states
$\left| {{\varphi _k}} \right\rangle  \propto \left| 0 \right\rangle  + {{\mathop{\rm e}\nolimits} ^{(2\pi iks/N)}}\left| 1 \right\rangle $

Step2 Give the tensor product of the above states 	
$$\sum\limits_{} {\exp (2\pi i({a_1}({x_1}) + {a_2}({x_2}) + ...{a_m}({x_m}))s/{2^n})} \left| {{x_1},{x_2},...,{x_m}} \right\rangle $$

Step3 Use the transition quantum state generating algorithm to get the transition quantum state
$\left| {{x^{(1)}}} \right\rangle  + {( - 1)^s}\left| {{x^{(2)}}} \right\rangle  + {( - 1)^{ts}}\left| {{x^{(3)}}} \right\rangle ...$

Step4 Call the algorithm SV and prepare the measurement basis$\left| {{x^{(i)}}} \right\rangle ,i \in \{ 1,2\} $, then measure the transition quantum state to get the target state
$$\left| 0 \right\rangle  + {( - 1)^s}\left| 1 \right\rangle $$

Step5 Get the parity of $s$ by the measurement in $\left|  \pm  \right\rangle $ basis.

\section{Algorithm analysis}
In this section, we analyze the algorithm in section 4 from the aspects of correctness, success probability, time complexity and space complexity.
\\\textbf{Correctness and success probability} The algorithm fails in the following situations: (1) in step 3 of the transition quantum state generating algorithm, the number of solution of $a(x) = c\bmod {2^{n - 1}}$ is less than 2; (2) in step 3 of the transition quantum state generating algorithm, after the eliminating global phase factor $\exp (2\pi ia({x^{(1)}})s/{2^n})$, the phases of $\left| {{x^{(1)}}} \right\rangle ,\left| {{x^{(2)}}} \right\rangle $ are 1, that is, we can¡¯t get the state $\left| {{x^{(1)}}} \right\rangle  + {( - 1)^s}\left| {{x^{(2)}}} \right\rangle  + ...$; (3) in section 4, algorithm SV is failed to get the right solution of low density subset sum problem.

For the situation (1), we give the theorem 6.1
\\\textbf{Theorem 5.1} for the equation ${a_1}{x_1} + {a_2}{x_2} + ... + {a_m}{x_m}\begin{array}{*{20}{c}}
{}
\end{array}\bmod {2^{n - 1}}$, where ${a_i}{x_i} = \sum\limits_{j = 1}^m {{a_{ij}}{x_{ij}}} $, denote $\tau $ the number of vectors$({x_1},...{x_m})$satisfy ${a_1}{x_1} + {a_2}{x_2} + ... + {a_m}{x_m}\begin{array}{*{20}{c}}
{}
\end{array}\bmod {2^{n - 1}} = c$, then the probability of $\tau  \ge 2$ is ${P_\tau } \approx \frac{3}{5}$.
\\\textbf{Proof}: as ${a_{ij}}$ is randomly chosen, for any $({x_1},...{x_m}) \in {\{ 0,1\} ^n}$, the probability of ${a_1}{x_1} + {a_2}{x_2} + ... + {a_m}{x_m} \equiv c\begin{array}{*{20}{c}}
{}
\end{array}\bmod {2^{n - 1}}$ is at least $\frac{1}{{{2^{n - 1}}}}$, ${P_1}$ denotes the probability that there is only one vector $({x_1},...{x_m})$ that satisfies the equation and ${P_0}$ denotes the probability that no vector $({x_1},...{x_m})$ satisfies the equation. Obviously,${P_\tau } = 1 - {P_1} - {P_0}$, and ${P_0} = {(\frac{{{2^{n - 1}} - 1}}{{{2^{n - 1}}}})^{{2^n}}}$,${P_1} = C_{{2^n}}^1(\frac{1}{{{2^{n - 1}}}}){(\frac{{{2^{n - 1}} - 1}}{{{2^{n - 1}}}})^{{2^n} - 1}}$, so
$$\begin{array}{l}
{P_\tau } = 1 - C_{{2^n}}^1(\frac{1}{{{2^{n - 1}}}}){(\frac{{{2^{n - 1}} - 1}}{{{2^{n - 1}}}})^{{2^n} - 1}} - {(\frac{{{2^{n - 1}} - 1}}{{{2^{n - 1}}}})^{{2^n}}}\\
 = 1 - {2^n}(\frac{1}{{{2^{n - 1}}}}){(1 - \frac{1}{{{2^{n - 1}}}})^{{2^{n - 1}} \cdot \frac{{{2^n} - 1}}{{{2^{n - 1}}}}}} - {(1 - \frac{1}{{{2^{n - 1}}}})^{{2^{n - 1}} \cdot 2}}
\end{array}$$

Since $\mathop {\lim }\limits_{n \to \infty } {(1 - \frac{1}{n})^n} = \frac{1}{e}$, then we get ${P_\tau } \approx 1 - 3{(\frac{1}{e})^2} \approx \frac{3}{5}$.

For the situation (2), if we obtain the solutions ${x^{(1)}},{x^{(2)}}$ of $a(x) \equiv c\bmod {2^{n - 1}}$, for some ${t_1},{t_2} \in Z$, there exist $a({x^{(1)}}) = {t_1}{2^n} + a'({x^{(1)}})$ and $a({x^{(2)}}) = {t_2}{2^n} + a'({x^{(2)}})$, after the measurement, the phase of the collapsed state is
$$\exp (2\pi ia({x^{(i)}})s/{2^n}) = \exp (2\pi i{t_i}s) \cdot \exp (2\pi ia'({x^{(i)}})s/{2^n}) = \exp (2\pi ia'({x^{(i)}})s/{2^n})$$

Since $a({x^{(1)}}) \equiv c\bmod {2^{n - 1}}$and $a({x^{(2)}}) \equiv c\bmod {2^{n - 1}}$, we know that
$$a'({x^{(1)}}) - a'({x^{(2)}}) \equiv 0\begin{array}{*{20}{c}}
{}
\end{array}\bmod {2^{n - 1}}$$

Thus, after eliminating global phase factor $\min \{ a'({x^{(1)}}),a'({x^{(2)}})\} $ of the collapsed state (without loss of generality we assume $\min \{ a'({x^{(1)}}),a'({x^{(2)}})\}  = a'({x^{(1)}})$), we can obtain the following quantum state$$\left| {{x^{(1)}}} \right\rangle  + {( - 1)^{ts}}\left| {{x^{(2)}}} \right\rangle  + ...$$

Where $t \in \{ 0,1\} $, the value of $t$ depends on the $n'th$ bit of $a'({x^{(1)}}),a'({x^{(2)}})$, since the quantum states are randomly chosen, the probability of $t = 1$ is $\frac{1}{2}$.

For the situation (3), the success probability of algorithm SV mainly depends on the quality of the reduced basis produced by LLL algorithm and the density of the subset sum problem. When $\sqrt n $ is small, it will be quiet efficient to find the shortest vector, when $\sqrt n $ is increased, the density will be decreased, from the previous test results, as $d \to 0$(that is,$\sqrt n  \to \infty $), the success rate of algorithm SV will be 1. However, these conjectures have not been proved rigorously yet.
\\\textbf{Time complexity} The running time of the algorithm is taken in the following aspects: (1) Preparing the states $\left| {{\varphi _k}} \right\rangle  \propto \left| 0 \right\rangle  + {{\mathop{\rm e}\nolimits} ^{(2\pi iks/N)}}\left| 1 \right\rangle $. It needs $O({n^3})$ quantum gate operations and $O(n)$ oracle calls. (2) Transition quantum state generating algorithm needs $O(\sqrt n )$ oracle calls. (3) Computing ${x^{(1)}},{x^{(2)}}$ needs solving $2{m^2}$ equations which leads $2{m^2}$ times calling of algorithm SV, the algorithm is not guaranteed to produce the shortest vector in polynomial time when $\sqrt n $ is large, but in practice it does, J. C. Lagarias and A.M.Odlyzko gave the experimental results that the running time of algorithm SV was about $O(m{(\log \max {a_{ij}})^3})$ over a wide range of values of $m$ and $\max {a_{ij}}$, up to $m = 80$, that is, for $n \le 6400$, we can solve the dihedral hidden subgroup problem in polynomial time. However, it seems hard to give the rigorous bound of time complexity for larger $n$.
\\\textbf{Space complexity} The algorithm needs $O(n)$ qubits to store the quantum state and $O({n^2})$ classical space to store the lattice basis and variables produced in LLL algorithm.

\section{Conclusions}
As large classical space is required in the precious algorithm for the dihedral hidden subgroup problem, a new quantum algorithm is designed. First, a transition quantum state generating algorithm is designed by constructing a phase filter function. Next, using the method of solving low density subset problem, the information of ground state ${x^{(i)}}$ is obtained by calling algorithm SV, then the measurement basis $\left| {{x^{(i)}}} \right\rangle $ are prepared and the transition quantum state is measured to get the target quantum state. Finally, the parity of $s$ is obtained by measurement in $\left|  \pm  \right\rangle $ basis. The algorithm decreases time and space complexity obviously by comparison with the results in \cite{13}.



\end{document}